# A High-Performance Cellular Automaton Model of Tumor Growth with Dynamcially Growing Domains

Jan Poleszczuk[1], Heiko Enderling[2]

[1]College of Inter-faculty Individual Studies in Mathematics and Natural Sciences, University of Warsaw, Warsaw Poland

[2]Integrated Mathematical Oncology, Moffitt Cancer Center and Research Institute, Tampa, FL, USA

Email: j.poleszczuk@mimuw.edu.pl, heiko.enderling@moffitt.org

**Abstract**

Tumor growth from a single transformed cancer cell up to a clinically apparent mass spans many spatial and temporal orders of magnitude. Implementation of cellular automata simulations of such tumor growth can be straightforward but computing performance often counterbalances simplicity. Computationally convenient simulation times can be achieved by choosing appropriate data structures, memory and cell handling as well as domain setup. We propose a cellular automaton model of tumor growth with a domain that expands dynamically as the tumor population increases. We discuss memory access, data structures and implementation techniques that yield high-performance multi-scale Monte Carlo simulations of tumor growth. We present simulation results of the tumor growth model and discuss tumor properties that favor the proposed high-performance design.

## Introduction

Simulating complex cellular automata is still a great challenge despite advances in computational power of modern computers in recent years. Cellular automata are increasingly used to simulate tumor growth dynamics [1-15]. Whilst many efficient ways exist to simulating deterministic and synchronous cellular automata such as Conway's 'Game of Life' [16], high-performance simulation of stochastic cancer cell kinetics and emerging multi-scale tumor population dynamics is still in is infancy. In Monte Carlo cancer models cells are not governed by simple deterministic rules but by coupled internal states and non-trivial interactions with the continuously changing local environment. Additionally, tumor population dynamics emerge from the interaction of millions of cells, and often the development of such populations from few initial cells needs to be simulated, which poses problems of bridging many temporal and spatial scales. Due to the stochastic nature of single cell kinetics many simulations need to be performed in order to obtain averaged and statistically significant results. To further complicate matters, in typical tumor growth models many parameters need to be estimated in high-dimensional parameter sweeps and sensitivity analyses need to be performed to study parameter influence on overall dynamics.

The main advantage of utilizing cellular automata in cancer modeling is the ability to formalize experimentally observable single-cell kinetics [17, 18] and observe emerging population level dynamics without a-priori knowledge of tumor behavior. Because of their apparent resemblance of *in vitro* cell culture models, cellular automata may be referred to as *in silico* experiments [19]. Automata simulations enable visualization, measurement and perturbation of cell kinetics as well as their interaction with the environment. Herein we describe a simple cellular automaton tumor growth model, and discuss computer memory access, data structures, domain setup and implementation techniques that enable high performance multi-scale simulations.

**Tumor Growth Model**

A cancer cell is an individual entity that occupies a single grid point of $(10\mu m)^2$ on a two-dimensional regular square lattice. Each cancer cell is characterized by its specific trait vector [cct, $\rho$, $\mu$, $\alpha$] denoting cell cycle time, remaining proliferation potential, migration potential and probability of spontaneous death, respectively [20]. We assume a heterogeneous tumor population consisting of so-called cancer stem cells and non-stem cancer cells. Cancer stem cells are assumed to be immortal and have unlimited proliferation potential (i.e., $\alpha=0$, $\rho=\infty$), whereas non-stem cancer cells can only divide a limited number of times $\rho_{max}$ before cell death. Each cell type can divide symmetrically to produce two daughter cells with parental phenotype. The populations are coupled through asymmetric division of cancer stem cells. Asymmetric division has a probability 1-$p_s$ (where $p_s$ is the probability of symmetric cancer stem cell division) to produce a cancer stem cell and a non-stem cancer cell with $\rho=\rho_{max}$, which decreases with each subsequent non-stem cell division (**Figure 1A**). Cells need adjacent space for migration and proliferation, and cells that are completely surrounded by other cells (eight on a two-dimensional lattice; Moore neighborhood) become quiescent (**Figure 1B**). In unsaturated environments, cells proliferate and migrate into vacant adjacent space at random. Cells can undergo spontaneous death independent of environment saturation with rate $\alpha$ and will be instantaneously removed from the system.

Time is advanced at discrete time intervals $\Delta t = 1/24$ day (i.e., 1 hour), and 24 simulation steps represent one day. At each simulation step, cells are considered in random order to minimize lattice geometry effects and the behavior of each cell is updated. Cell proliferation and migration are random events with the respective probabilities scaled to the simulation time step. Cell proliferation and migration are temporally mutually exclusive events. We assume that at each simulation step, if there is no spontaneous death event, cells proliferate with probability $p_d =$(24hours/cct)$\times\Delta t$, migrate with probability $(1-p_d)p_m$ and die with probability $\alpha$. Let $p_m= \mu\times\Delta t$ where $\mu$ denotes cancer cell motility rate.

**Implementation**

**Memory architecture and data access**

High-performance simulations require fast access to allocated memory and cached data. How memory is handled depends heavily on simulation design and used data structures and procedures. The memory in modern desktop PCs has three layers: the built-in cache memory has the fastest access time (1-20 ns) but a very limited size; random access memory (RAM) is slower (50-100 ns) but much larger; and hard disk drives (HDD) whilst having large memory have the slowest access time (5-10 ms) (**Figure 2**).

State-of-the-art processors may have up to 24 Megabytes (MB) of cache memory and can address up to 4096 Gigabytes (GB) of RAM (c.f., Intel® Xeon® E7-8830). Cache memory stores the most frequently used RAM locations to reduce access time to necessary information [21]. Due to limited memory size, cached content constantly changes throughout simulations. Simulation time decreases when the *spatial locality* property is unsatisfied, i.e. the CPU frequently requires access to data elements that are in distant storage locations, and hence not stored simultaneously in cache memory. If a so-called *cache miss* occurs data needs to be retrieved from much slower RAM or even HDD memory. High frequencies of *cache misses* dramatically reduce computation speed, so optimized algorithms should minimize *cache miss* events.

Let us consider a two-dimensional rectangular lattice coded by a two-dimensional array as commonly used in cellular automata. As computer memory is arranged linearly, higher-dimensional arrays are stored row after row (or column after column). Therefore, as an array element is accessed only parts of its immediate spatial neighborhood will be stored in the cache. Especially for large lattices, 2 of 4 neighbors (2-D von Neumann neighborhood) or 6 of 8 neighbors (2-D Moore neighborhood) are *cache missed*. Whilst convenient at implementation, access to all cell neighbors in two- and three-dimensional arrays is memory inefficient and slow.

**Population geometry and data type optimization**

Which data structures are best to use depends on the cellular process that are considered as well as the geometry of the emerging population. Prostate tumors, for example, have a very dense, compact structure whereas glioblastoma brain tumors are highly diffusive. Such density difference may be represented by the number of cells on the computational lattice per area or volume. Let us define a dense tumor as a population of cells where each lattice point is occupied by a cell with probability p=0.99 (i.e., 99%) and a diffusive tumor occupies lattice points with p=0.5. For cells to migrate or proliferate adjacent lattice points need to be vacant. The most efficient data structure for obtaining vacant neighbor lattice sites will be dependent on expected tumor density - either many or few neighbors for most cells. To determine cell neighborhood vacancies, a simple array keeping *boolean* information about lattice points occupied by cells will be highly inefficient for dense tumors. Let us consider morphological erosion, where each cell is removed from the lattice if it is not completely surrounded by other cells. For dense population geometries, a coded array containing information about number of vacant spots in the cell neighborhood may be more efficient to avoid unsuccessful scanning of each neighboring lattice point for vacancy (**Figure 3A,B**). Appropriate use of C++ data type *char* will not introduce a memory tradeoff as both *char* and *boolean* require one byte of memory. Using intuitive *int* instead of *char* will require four times more memory and increases computation time as less cached memory is available. A computationally expensive draw-back of a coded lattice is the requirement to update all neighboring lattice codes when occupancy of a single grid point changes, which makes this approach less efficient for diffusive tumors (**Figure 3C**).

**Random neighbor selection**

Monte Carlo tumor growth simulations frequently require obtaining a free neighboring lattice site at random, for example for migration or proliferation. A naïve approach may consider all neighboring lattice sites, store those that are vacant in a temporary vector, and finally select a vector element at random. Alternatively, neighboring lattice sites may be accessed in random order and the first encountered vacant position is selected (**Figure 4A**). This simple alternative random access method significantly decreases simulation time in dense (**Figure 4B**) and diffusive tumors (**Figure 4C**) with increasing lattice size. While the naïve procedure is much slower for diffusive tumors because more vacant lattice sites have to be stored in temporary vectors, the alternative random access approach performs equally well irrespective of tumor type. Dependent on the modeled cellular processes, additional alterations or improvements may be required. For example, one may choose to store hashed information about the cell neighborhood in the lattice. In particular, a limited number of possible cell neighborhood configurations may be encoded in identifying keys.

**Random ordering**

Many programming languages provide efficient procedures and data structures that can be utilized for cellular automata design in combination with simulation specific code. In asynchronous stochastic cellular automata of tumor growth, random cell ordering and random access of cells is fundamental. A naïve implementation of selecting cells in a random order may consist of

1. From a vector containing all cells, pick a cell at random by drawing a random positive integer not larger than the vector length.
2. Erase the selected cell from the vector to avoid its reselecting.
3. Repeat steps 1 and 2 until there are no cells left in the vector.

The C++ Standard Template Library (STL) provides numerous algorithms to perform search, sort and shuffle operations. Random shuffle rearranges all elements in a specified range randomly in a single invocation. The STL random shuffle procedure reduces computation time compared to the naïve approach by multiple orders of magnitude even for small vector sizes (**Figure 5**), clearly demonstrating the power and importance of using standard language-specific data structures and algorithms for high-performance simulations.

**Dynamically growing domains**

To simulate a growing tumor population from a single cancer cell computational lattice-induced boundary constraints need to be avoided. An appropriate lattice size must be selected dependent on the achievable tumor size, which requires a priori knowledge about emerging population dynamics, tumor density and cell diffusibility. A dense radially symmetrically growing two-dimensional tumor population of 100,000 cells could well fit into a 400x400 lattice. Cells in a highly diffusive irregular tumor, however, will likely hit the boundary of such lattice early during tumor growth. Whilst a sufficiently large lattice could ensure avoidance of boundary contact, memory requirements and computing performance limit such approach. Large amounts of computational resources would be wasted especially early in population expansion when only a few cells are present.

One possibility is to use dynamic data structures such as a C++ standard template library (STL) map, which can be understood as an associative container that stores elements formed by the combination of a key value (i.e., position on the lattice) and a mapped value (i.e., occupancy). Unfortunately, accessing elements in a map is logarithmic in size, which for large tumor sizes dramatically decreases computational performance (**Figure 6**). We propose a dynamically allocated array with associated procedures that expand the lattice upon cell boundary contact by a fixed amount of lattice points. While a static large array of 1000x1000 lattice points is the most efficient for large tumors it is inefficient for small tumors up to 10,000 cells due to unoccupied lattice sites occupying large amounts of memory. The dynamically expanding array is most efficient for small tumors. When the tumor population approaches carrying capacity of a static lattice, both static and dynamic lattice perform similarly.

**Tumor growth simulations**

Let us initialize tumor growth simulations with one cancer stem cell located in the center of a square lattice with trait vector [cct=24hours, $\rho_{max}$=10, $\mu$=100μm/day, $\alpha$=1%] and $p_s$=0.1. These parameter values have previously been shown to enable fast dense tumor growth [20]. Tumor growth dynamics with other parameters have been discussed elsewhere [22-24]. We simulate tumor growth for t=180 days using an intuitive implementation (naïve code) and compare to an implementation with a combination of above-discussed improvements (improved code). The naïve simulation is executed on a fixed 750x750 square lattice, whereas the improved simulation is initiated on a 50x50 square lattice with dynamically expanding domains. Due to the stochastic nature of the model we simulate N=100 (improved) and N=77 (naïve) independent tumors and report average results. Both implementations yield similar population sizes with comparable cancer stem cell and non-stem cancer cell numbers (**Figure 7**). While the naïve code executes in an average of 4212 seconds (>70 minutes) the improved code executes in 51 seconds (<1 minute) – an 82-fold reduction in computing time. The high-performance of the improved code is due to the dynamically expanding domain as well as efficient access to information on vacant neighboring lattice sites. More than 70% of all cells at the final time point of the simulation have no adjacent space, and less than 5% of cells have two or more vacant lattice sites to migrate or proliferate into (**Figure 7**). Graphical visualization of tumor morphologies at different time points show that tumors simulated with either implementation technique are non-differentiable beyond intrinsic stochastic effects (**Figure 8**).

**Discussion**

Cellular automata are frequently used to simulate solid tumor growth and cancer stem cell dynamics

[1,25,26]. Intuitive computer implementation of stochastic cellular automata for tumor modeling is counterbalanced by its performance. Although cellular automata are lattice-based, naïvely implementated *boolean* arrays may lack computational efficiency. We set out to compare C++ data structures, memory-efficient procedures and dynamic domains to decrease computing time. As extension to three spatial dimensions is trivial, we presented implementation details in two dimensions for clarity. We found that simple substitutions in intuitive cellular automaton implementations significantly decrease computing time. First, appropriate use of data type *char* over *int* provides a 4-fold reduction in memory allocation. Second, consideration of a coded lattice that holds information about a cell's neighborhood vacancies rather than *boolean* information whether or not a cell is occupying that lattice point significantly decreases computation time if queries about adjacent space are frequently required for cell decisions. Third, utilization of the C++ STL shuffle method to provide a random order of elements proves superior to repeatedly selecting single elements at random positions within a vector. Finally, we presented a dynamically growing domain that evolves according to the population size, which keeps compuation time exceptionally low compared to large lattices when the population is small. When all of these adjustments are combined into a simulation of cancer stem cell-driven solid tumor growth, the improved implementation yields a high-performance over the naïve approach. In simulations of tumor growth for 180 days from a single cell to a populaiton of about 140,000 cells, the presented high-performance cellular automaton yields an 82-fold reduction in computing time while reproducing the results of the naïve implementation. We believe the developed high-performance cellular automaton will serve as a template for future simulations of solid tumor growth as well as other population dynamics models. We share the source code for the presented naïve and improved code on our personal websites and the `sourceforge.net` repository.

## Acknowledgements

This work was partially supported by the European Social Fund, contract number UDA-POKL. 04.01.01-00-072/09-00 (J.P.) and the NIH/NCI Integrative Cancer Biology Program (5U54 CA113007) (H.E.).

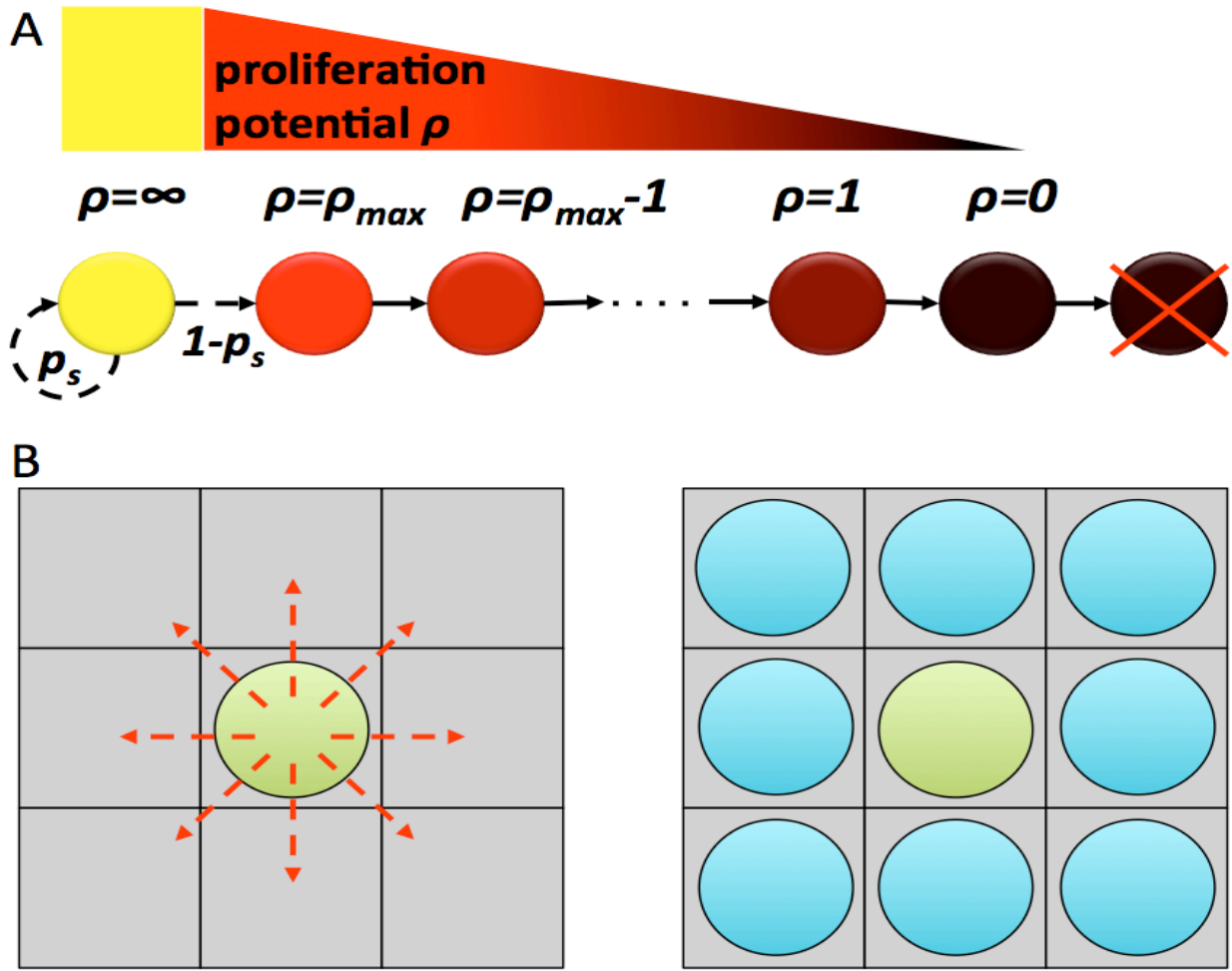

**Figure 1**. A) Cancer cell lineage; yellow: cancer stem cell, red-black: non-stem cancer cells with decreasing potential. B) Tumor cells populate the computational lattice by cell migration and cell proliferation. A cell can randomly migrate to or place a daughter cell into one of the eight adjacent lattice points subject to availability. A cell becomes quiescent if all adjacent lattice points are occupied.

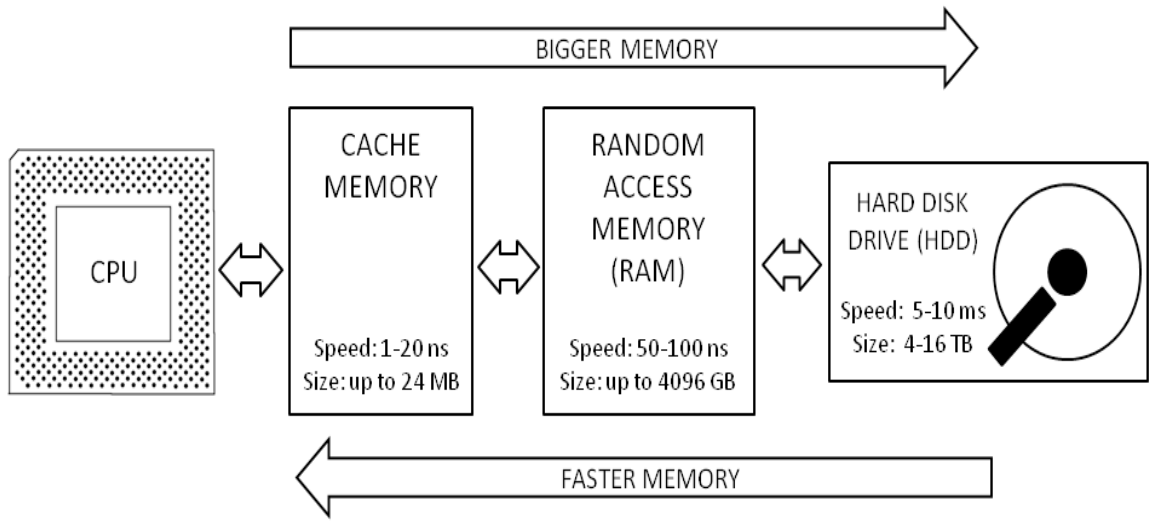

**Figure 2.** Typical architecture of modern desktop PCs. Central processing unit (CPU) reads memory directly from the fastest cache, which, if the data is unavailable reads from the slower random access memory (RAM) and if needed from the biggest but slowest hard disk drive (HDD).

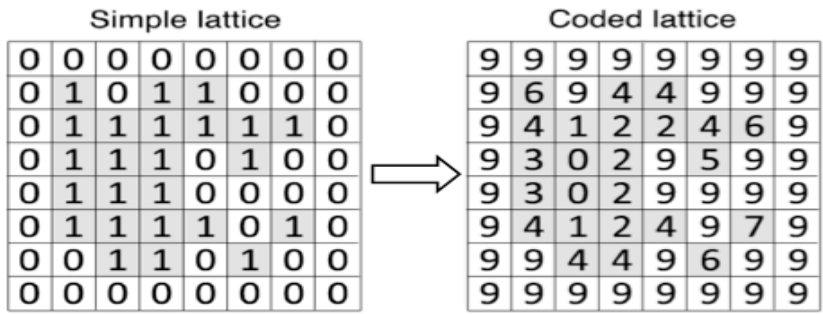

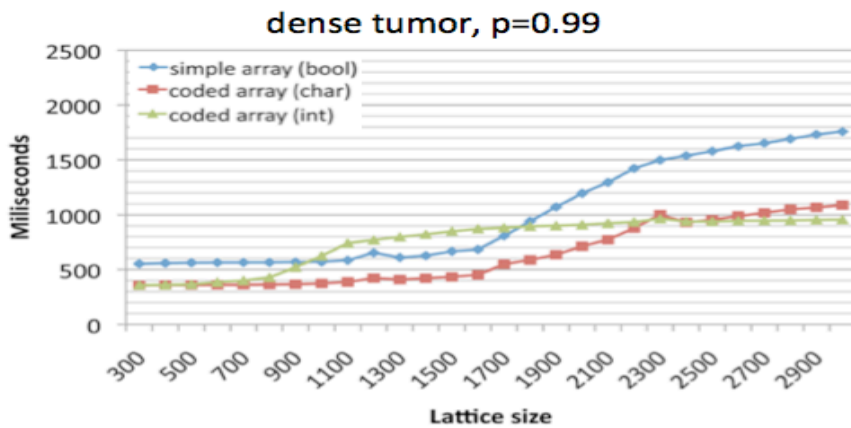

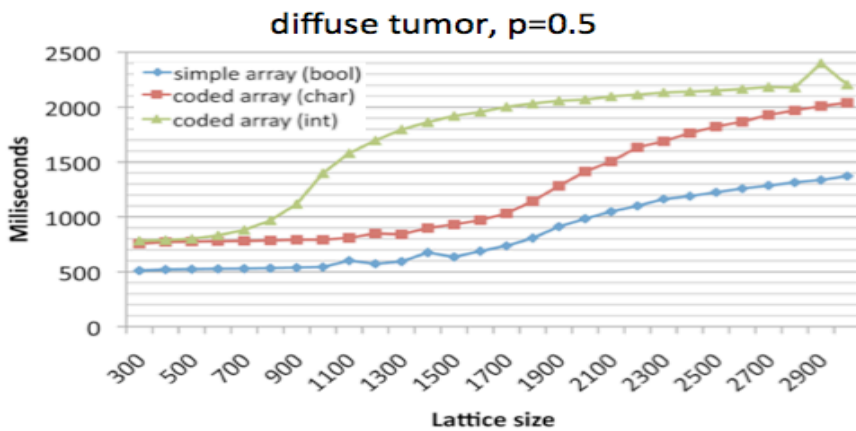

**Figure 3**. Morphological erosion on a simple *Boolean* array and a coded lattice. A) Illustration of transformation from simple *Boolean* lattice into coded lattice. In a coded lattice each occupied grid point contains the information about the number of free spots in its neighborhood, and 9 represents an empty grid point. B and C) Comparison of the evaluation times for dense tumors (B) and diffuse tumors (C). Three considered data structures are simple *Boolean* array (blue diamonds) and coded lattices using *char* (red squares) or *integer* values (green triangles).

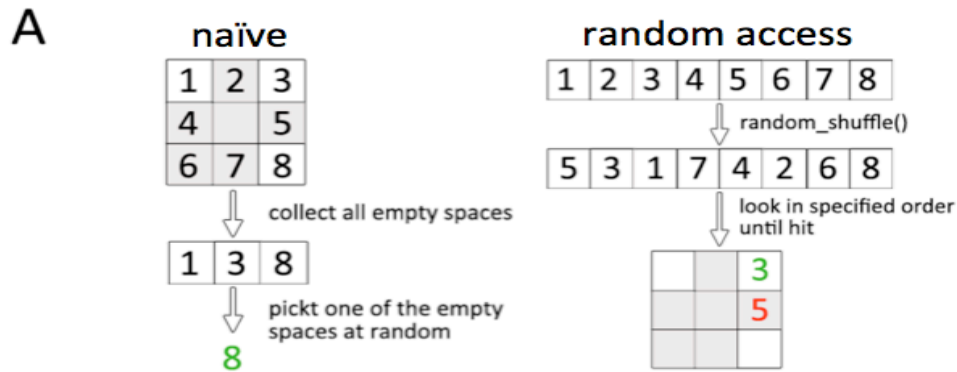

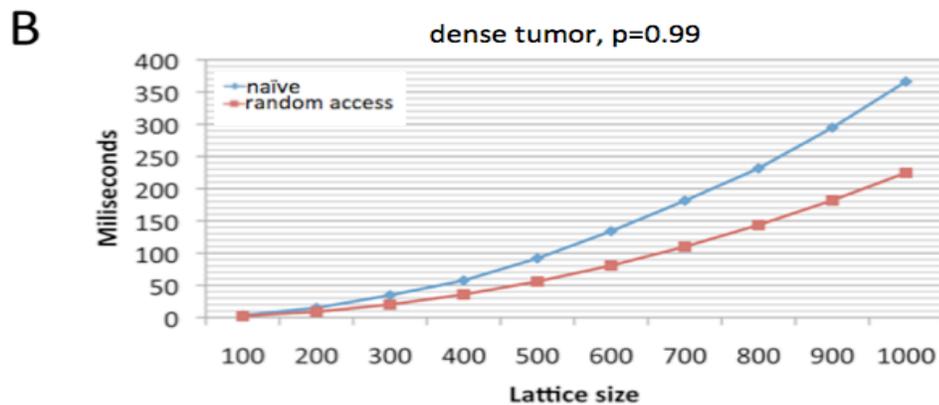

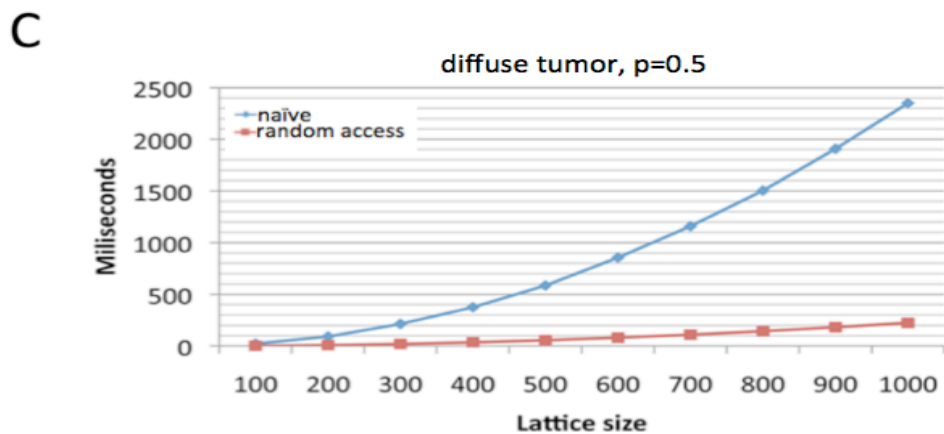

**Figure 4.** Comparison of two different procedures to select a random free spot from cell neighborhood. A) Naïve procedure visits all neighboring spots, temporarily stores those that are vacant and choses a random element from the temporary vector. Random access procedure uses a predefined vector of neighboring lattice sites that is randomly shuffled. The cell neighborhood is searched in that random order and returns first visited vacant site. B and C) Average evaluation times of different lattice sizes for dense tumors (B) and diffusive tumors (C) for several thousand iterations of the procedures. Error bars omitted for clarity.

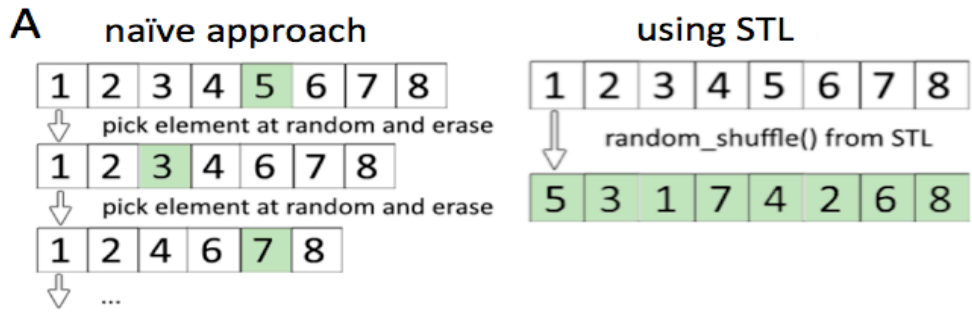

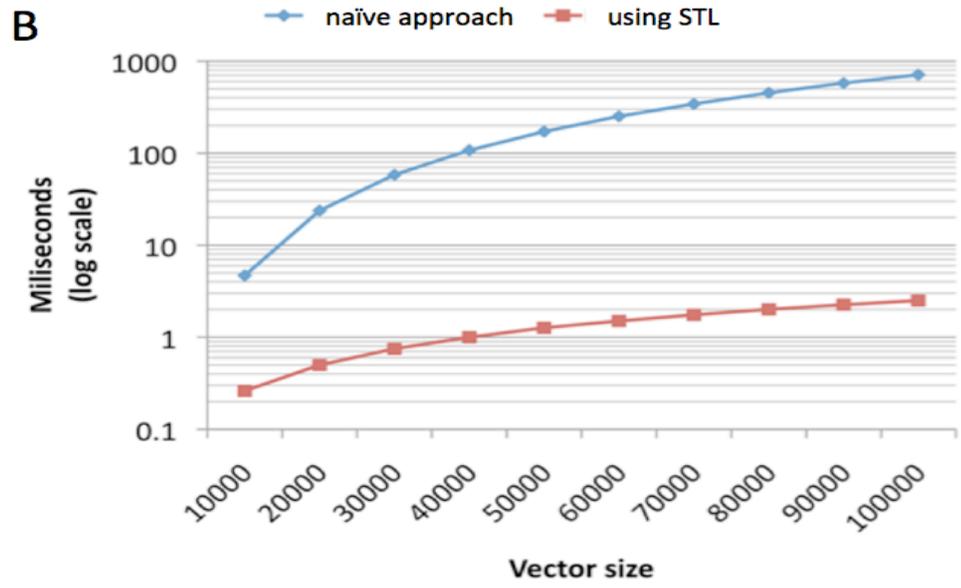

**Figure 5.** Comparison of the efficiency of the built-in STL shuffling procedure and the procedure coded in the naïve way. A) Naive procedure iterates the steps consisted of picking a random element from the vector and then erasing it until the initial vector is empty. Single invocation to the STL procedure gives a shuffled vector to determine order. B) Evaluation times for both procedures for different sizes of the shuffled vector.

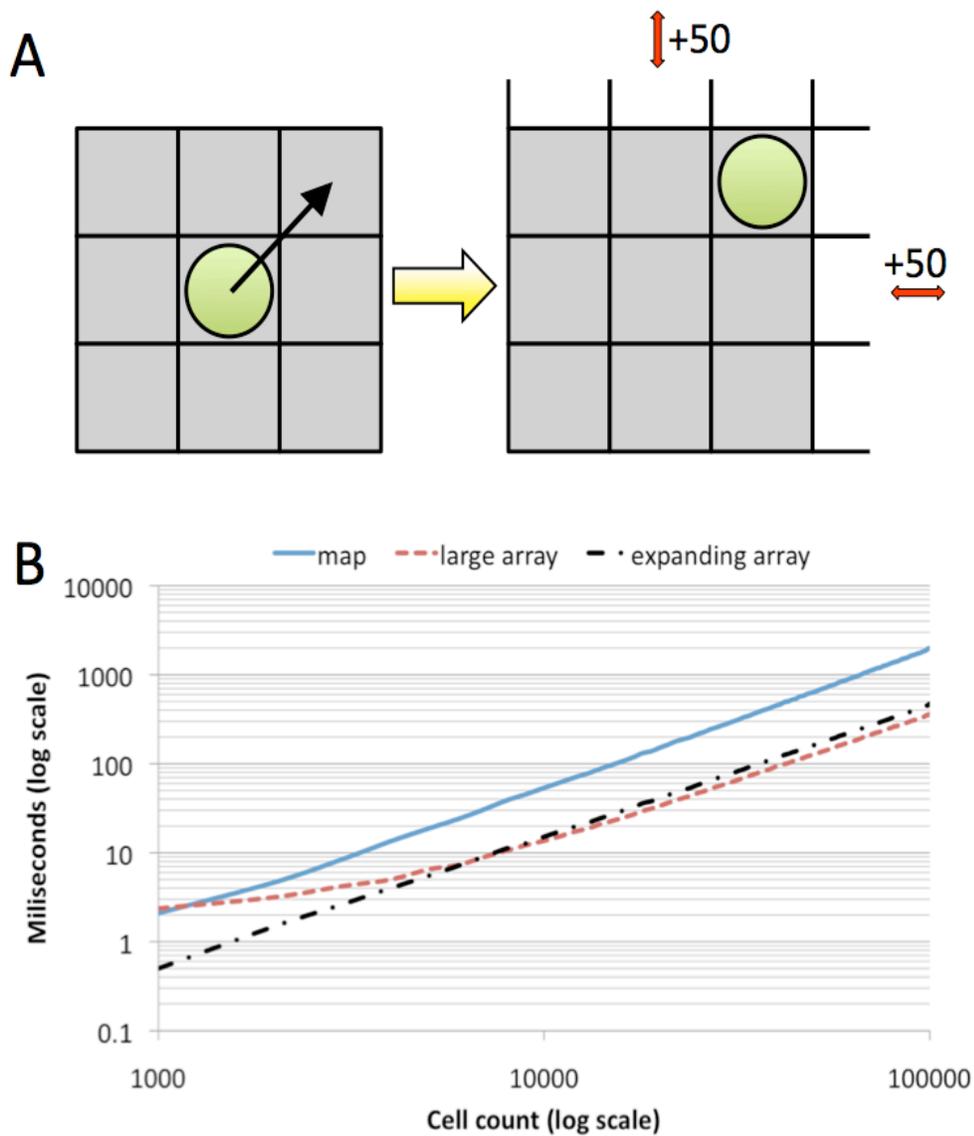

**Figure 6.** Dynamically growing domain. A) Computational lattice expands by 50 grid points in the direction(s) where a cell reaches the boundary. B) Simulations of tumor growth from a single cell when each cell proliferates into a random free adjacent spot until the tumor population reaches the predefined final size. Evaluation of computing time to reach different cell counts for three types of domains: STL map (solid blue curve), large array (red dashed) and expending array (black dash-dotted).

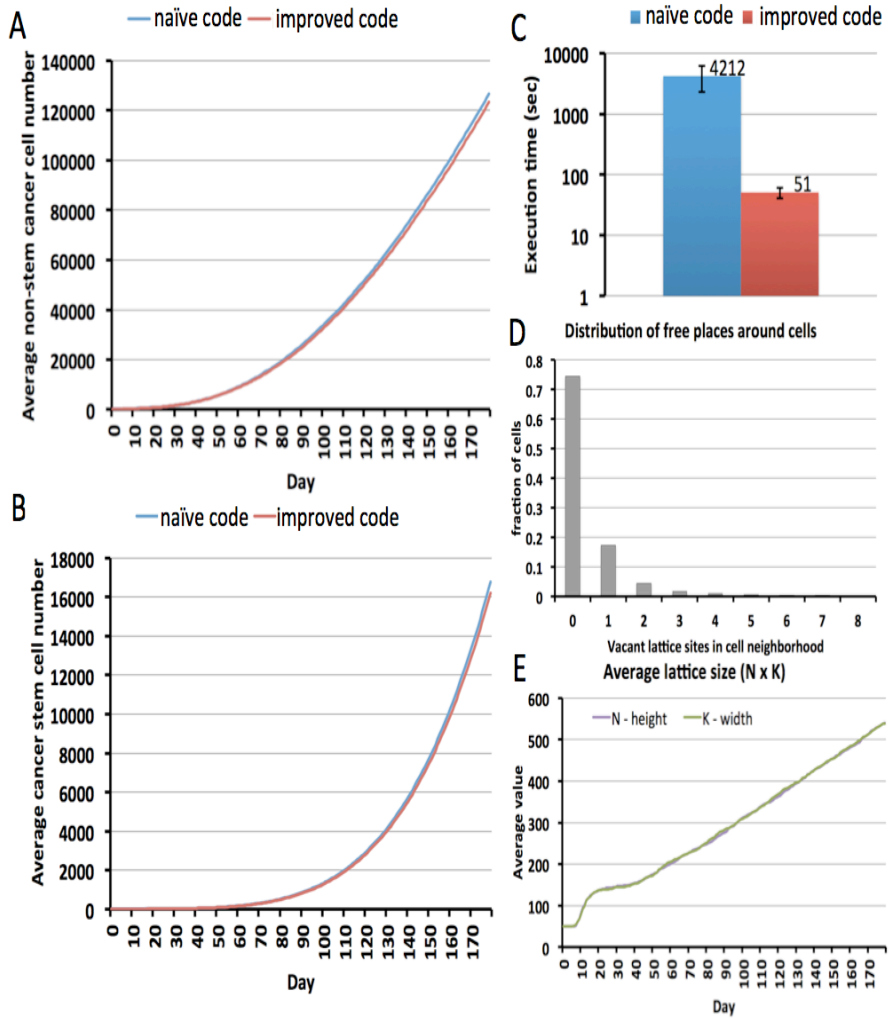

**Figure 7.** Model results for simulating 180 days of tumor growth initiated by one cancer stem cell. Naïve (blue curve) and improved (red) code simulate comparable tumor growth dynamics with similar non-stem cancer cell (A) and cancer stem cell (B) numbers. C) Average simulation times and standard deviations. D) Distribution of vacant lattice sites in cell neighborhood at final simulation time point. E) Evolution of average lattice size in the improved code. Initial lattice size 50x50, final lattice size 550x550. Shown are averages (standard deviations ommited in most panels for clarity) for N=100 (improved code) and N=77 (naïve ) independent simulations.

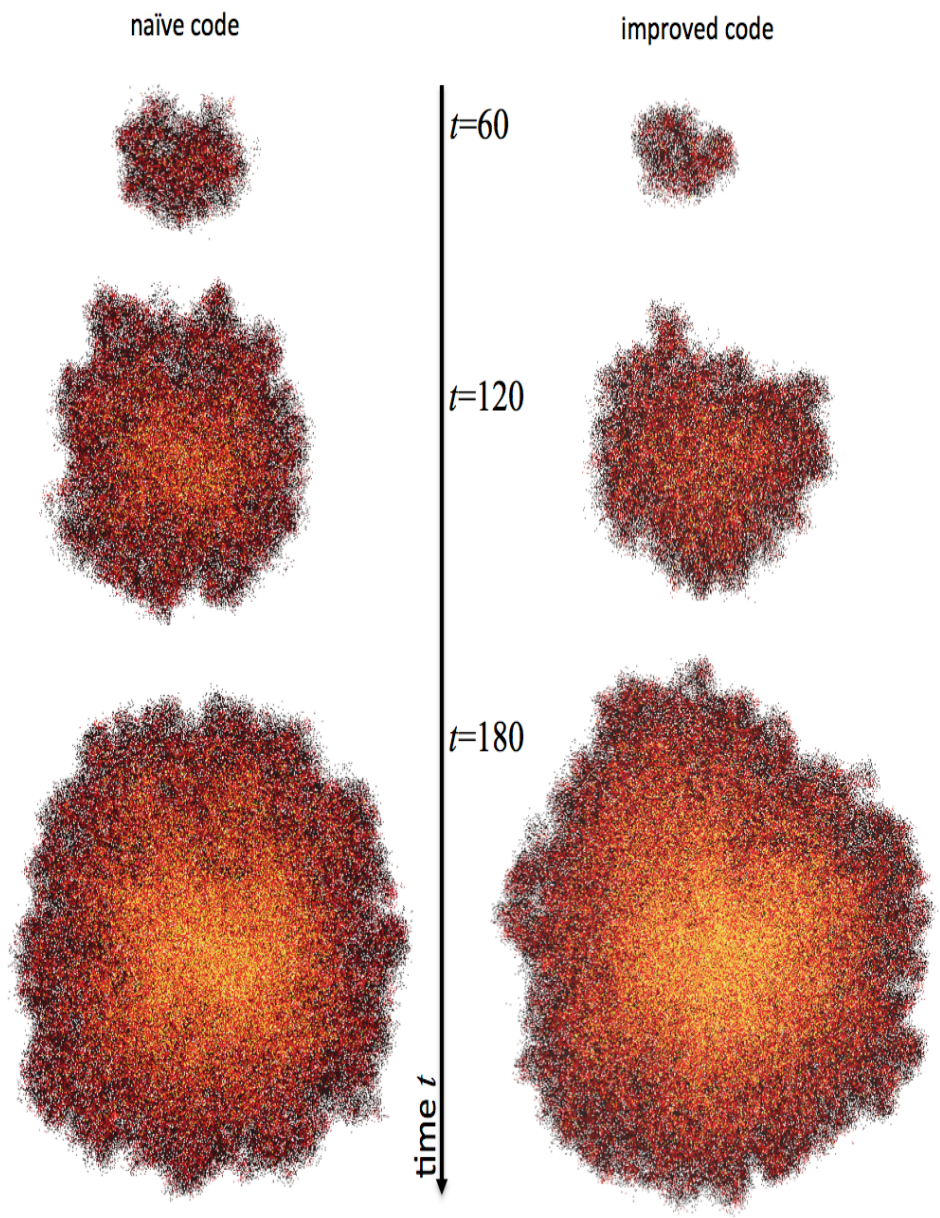

**Figure 8.** Representative tumor morphologies simulated with naïve (left) and improved (right) code. *Colors as defined in Figure 1A.*